\documentclass{emulateapj}
\usepackage{natbib,bm}
\usepackage{graphicx}
\usepackage{color}

\usepackage{mathptmx}
\usepackage{amsmath}

\newcommand{\be}{\begin{equation}}
\newcommand{\ee}{\end{equation}}
\newcommand{\bea}{\begin{eqnarray}}
\newcommand{\eea}{\end{eqnarray}}
\newcommand{\ba}{\begin{align}}
\newcommand{\ea}{\end{align}}

\begin{document}
%\preprint{NORDITA-2015-42}
\title[Two-component superfluid hydrodynamics  of neutron star cores]{Two-component superfluid hydrodynamics  of neutron star cores}
\author{D. N. Kobyakov\altaffilmark{1}}
\affil{
Institute of Applied Physics of the Russian Academy of Sciences, 603950 Nizhny Novgorod, Russia\\
Institut f\"ur Kernphysik, Technische Universit\"at Darmstadt, 64289 Darmstadt, Germany
\\Department of Physics, Ume{\aa} University, SE-901 87 Ume{\aa}, Sweden
}
\author{C. J. Pethick\altaffilmark{2}}
\affil{The Niels Bohr International Academy, The Niels Bohr Institute, University of Copenhagen, Blegdamsvej 17, DK-2100 Copenhagen \O, Denmark\\
NORDITA, KTH Royal Institute of Technology and Stockholm University, Roslagstullsbacken 23, SE-106 91 Stockholm, Sweden}
\altaffiltext{1}{E-mail: dmitry.kobyakov@appl.sci-nnov.ru}
\altaffiltext{2}{E-mail: pethick@nbi.dk}
\date{}
\label{firstpage}

\begin{abstract}
We consider the hydrodynamics of the outer core of a neutron star under conditions when both neutrons and protons are superfluid.
Starting from  the equation of motion for the phases of the wave functions of the condensates of neutron pairs and proton pairs  we derive the generalization of the Euler equation for a one-component fluid.
These equations are supplemented by
the conditions for conservation of neutron number and proton number.
Of particular interest is the effect of entrainment, the fact that the current of one nucleon species
depends on the momenta per nucleon of both condensates.
We find that the nonlinear terms in the Euler-like equation contain contributions that have not always been taken into account in previous applications of superfluid hydrodynamics.
We apply the formalism to determine the frequency of oscillations about a state with stationary condensates and states  with a spatially uniform counterflow of neutrons and protons.
The velocities of the coupled sound-like modes of neutrons and protons are calculated from properties of uniform neutron star matter evaluated on the basis of chiral effective field theory.  We also derive the condition for the two-stream instability to occur.

\end{abstract}


\section{Introduction}

The outer core of a neutron star consists of a uniform fluid of neutrons, protons and electrons, with possibly other minority constituents.  The hydrodynamics of the core of a neutron star is important for studies of a variety of phenomena, among them stellar oscillations \citep{Mendell1991,  LindblomMendell1994,LindblomMendell2000,AnderssonComer2001},  collective modes of matter \citep{Epstein1988, BedaqueReddy2014}, as well as theories of spin-down and glitches in the rotation rate of neutron stars (For a review, see \citet{HaskellMelatos}). From microscopic calculations, protons are expected to be superconducting in the outer core, while the situation for neutrons is less clear because of the difficulty of calculating superfluid gaps at such densities with confidence.  In this paper we shall consider the case when the protons are superconducting and the neutrons superfluid.

The purpose of this paper is to derive the equations governing the long-wavelength, low-frequency behavior of the system.   We shall assume that thermal effects may be neglected: typical temperatures in neutron stars are of order 10$^8$K or \mbox{10 keV}, which is small compared with the Fermi energies of the components, which are of order 100 MeV.  We shall further assume that the superfluid gaps are large compared with the thermal energy $k_BT$, where $k_B$ is the Boltzmann constant and $T$ the temperature.   The basic variables in the approach we shall adopt are the density of neutrons, the density of protons, and the phases of the condensate wave functions of pairs of neutrons and pairs of protons.   This leads naturally to a description of the dynamics in terms of the gradients of the phases, which correspond to the momentum per particle of the condensates.  We shall show that this approach leads straightforwardly to equations for the dynamics, including nonlinear terms{, which agree with the work of \citet{Mendell1991}.}

Of particular interest in this paper is the influence of entrainment, the fact that there is a coupling between the currents of the two components.   To make the exposition as clear as possible, we shall derive the equations of motion by pedestrian methods.   We shall then show how they may be obtained from a Hamiltonian approach that exploits the fact that the phase of the condensate wave function of a component is the canonically conjugate variable to the density of that component \citep{LL1980}.
A particular focus of the work is to generalize the Euler equation for a one-component fluid to a two-component system, and we shall show that, in the Euler equations, there are contributions in the nonlinear terms in the Euler-like equations that have not always been considered in past applications, although they are implicit in the basic formalism (see, e,g, \citet{Mendell1991}).  These arise because the quantity determining the degree of entrainment is a function of the densities of the two components.
A preliminary report of many of the results in this article was given by \citet{Kobyakov_et_al2015}.

This article is arranged as follows.  The basic formalism is described in Section 2, where we work in terms of the phases of the wave functions for the neutron and proton pair condensates, and the neutron and proton number densities.  The equations of motion for the phases are described by a Josephson equation and  that of the nucleon densities by continuity equations.  Because of entrainment, the neutron number current depends not only on the gradient of the phase of the neutron condensate but also on the phase of the proton condensate, and similarly for the proton current.  In addition, entrainment affects the chemical potentials of nucleons.  The specific form of the Euler-like equations for the momentum per particle of the condensates is derived in Section 3.   Collective modes of oscillation about an initial situation where the condensates are stationary and the densities uniform are considered in Section 4.  There we also small deviations from a state with a uniform counterflow of neutrons and protons.  Applications to the outer core of a neutron star are described in Section 5, where we calculate collective mode velocities.  Section 6 contains a general discussion of, among other things, the relationship between our work and some earlier work on superfluid hydrodynamics.

\section{Basic formalism}

We shall consider long-wavelength, low-frequency phenomena, in which local charge neutrality is maintained and electrical currents are absent.
This is a good approximation for frequencies small compared with the electron plasma frequency and for wavelengths long compared with the Debye screening length for electrons.
Moreover, the hydrodynamic approximation implies that the frequency is smaller than the inverse of the electron relaxation time due to electron--electron collisions.
We shall also neglect dissipation due to Landau damping of the electron motion, which will be treated elsewhere \citep{Kobyakov_et_al2016}.
Under these conditions, the system behaves as a two-component system, one component being the neutrons and the other the protons and electrons. Throughout we shall work in an inertial frame of reference, and therefore the centrifugal and Coriolis forces will not appear explicitly.  We denote the phase of the superfluid order parameter for neutrons by $2\phi_n$ and that for protons by $2\phi_p$. To first order in the gradients of the phases, one may write the number current density of neutrons as
\be
{\bf j}_n=   \frac{n_{nn}}{m} {\bf p}_n + \frac{n_{np}}{m} {\bf p}_p
\label{current_n}
\ee
and that for protons as
\be
{\bf j}_p=   \frac{n_{pp}}{m} {\bf p}_p + \frac{n_{pn}}{m} {\bf p}_n,
\label{current_p}
\ee
where ${\bf p}_\alpha=\hbar \bm\nabla\phi_\alpha$ is the momentum per particle associated with the condensate and the response functions $n_{\alpha\beta}, \{\alpha,\beta=n,p\}$ generally depend on the density of neutrons, $n_n$, and the density of protons, $n_p$, but are independent of the gradients of the phases. Which mass is inserted in these equations is arbitrary, but the choice of the nucleon mass $m$ makes for simple expressions later in the analysis.  To avoid inessential complications, we shall neglect the difference between the neutron and proton masses.
The quantity $n_{\alpha \beta}/m$ is the long wavelength limit of the zero frequency neutron-current-density--proton-current-density response function and it is symmetrical in the indices $\alpha$ and $\beta$.

We shall assume that characteristic times for weak interaction processes are long compared with the timescales of the motions, and therefore the numbers of neutrons and of protons are separately conserved.
The continuity equation  is therefore
\be
\frac{\partial n_n}{\partial t}+\bm{\nabla}\bm{\cdot}\left( \frac{n_{nn}}{m} {\bf p}_n + \frac{n_{np}}{m} {\bf p}_p\right)=0
\ee
for neutrons and
\be
\frac{\partial n_p}{\partial t}+\bm{\nabla}\bm{\cdot}\left( \frac{n_{pp}}{m} {\bf p}_p +\frac{n_{np}}{m} {\bf p}_n\right)=0
\ee
for protons.
A separate continuity equation for electrons is not required, since the electron number density and current density are the same as those of the protons.

We are interested in situations where spatial variations are slow.  To determine how the phase of a state varies in time, we may therefore consider states in which the densities of neutrons and protons are uniform, and the gradients of the phases are uniform.  The equations of motion for the phases may be obtained by making use of the fact that in a state with energy $\mathcal E$ the  wave function varies in time as ${\rm e}^{-{\rm i}{\mathcal E}t/\hbar}$.  In terms of the ground states $\left|N_n, ~N_p\right>$ of the system with $N_p$ protons and $N_n$ neutrons, the superfluid order parameter for neutrons is
\bea
&&\left<N_n-2,~ N_p|  \psi_{n\uparrow}({\bf r}) \psi_{n\downarrow}({\bf r})|N_n,~ N_p \right> \nonumber \\
&&\hspace{8em}\sim {\rm e}^{-{\rm i}[\mathcal {E}(N_n, ~N_p)-{\mathcal E}(N_n-2,~ N_p)]t/\hbar},
\label{Schroedinger}
\eea
where $\psi_{n\sigma}({\bf r})$ is the annihilation operator for a neutron of spin $\sigma$.\footnote{For simplicity we consider the case of an S-wave superfluid, where pairing is in a spin singlet state.  For superfluids with anisotropic gaps the pairing amplitude must be defined for particles with specified momenta. }  We remark that the energies of the states are also functions of the gradients of phases of the superfluid order parameters.  The quantity ${\mathcal E}(N_n,~N_p)-{\mathcal E}(N_n-2,~N_p)$ is twice the neutron chemical potential including the contribution due to motion of the components, which we denote by $\mu_n^{\rm tot}$.\footnote{Since the phase is proportional to the difference of the energies of ground states whose neutron numbers differ by 2, it is a smooth function of $N_n$ and does not depend on whether $N_n$ is odd or even. }   From Eq.~(\ref{Schroedinger}) we conclude that
\be
\hbar \frac{\partial \phi_\alpha}{\partial t}=-\mu_\alpha^{\rm tot},
\ee
which is essentially Josephson's equation (see,~e.g., \citet{Varaquaux2015}).
In this article, we shall include in the calculations terms of second  order in  ${\bf p}_\alpha$, and therefore we need  the Hamiltonian to this order.  In the Hamiltonian formalism, the current density is given by
\be
{\bf j}_\alpha=\frac{\delta {\cal H}}{\delta {\bf p}_\alpha},
\ee
where ${\cal H}=\int d^3 {\bf r} ~{H}$ is the Hamiltonian, $H$ being the Hamiltonian density.

It follows from Eqs.\ (\ref{current_n}) and (\ref{current_p}) that the ``kinetic''{\footnote{We shall refer to as ``kinetic'' all contributions due to the motion of the components, including that due to entrainment.}} contribution to the Hamiltonian density is\footnote{In the literature, the symbol $\bf v$ is  used to denote an average velocity in some places and the momentum per unit mass of the condensate particles in others.  To avoid confusion,  we shall generally work with $\bf p$, the momentum per particle in the condensate.}
\bea
{H}^{\rm kin}&=&\frac{n_{nn}}{2m} p_n^2+\frac{n_{pp}}{2m}p_p^2 +\frac{n_{np}}{m}{\bf p}_n\bm{\cdot}{\bf p}_p \label{H^kin}\\
&=&\frac{( n_{nn}+n_{np})}{2m} p_n^2+\frac{(n_{pp}+n_{pn})}{2m} p_p^2 -\frac{n_{np}}{2m}({\bf p}_n-{\bf p}_p)^2.\nonumber \\
\label{H^kin2}
\eea
The final term in Eq.\ (\ref{H^kin2}) represents the effects of entrainment of the motions of the two fluids.

In the hydrodynamic description of a one-component fluid, the quantity $\Phi=\hbar \phi/m$ is commonly referred to as the {\it velocity} potential, since the fluid velocity is $\bm{\nabla} \Phi$.  However, we see from the considerations above that in multi-component systems, the phases $\phi_\alpha$ are more properly regarded as {\it momentum} potentials, since the momentum per particle of species $\alpha$ in the condensate is $\hbar \bm{\nabla} \phi_\alpha$.

The remaining contribution to the Hamiltonian density is the energy density of the system in the absence of gradients of the phases, which we denote by $E(n_n,n_p)$.  Thus the Hamiltonian density is
\be
{H}=E(n_n,n_p)+{H}^{\rm kin}.
\ee
The equations of motion for the phases are therefore
\be
\hbar\frac{\partial \phi_n}{\partial t}= - \frac{\delta {\cal H}}{\delta n_n} =-\frac{\partial H}{\partial n_n}=-\mu_n-\frac{\partial {H}^{\rm kin}}{\partial n_n},
\label{josephson_n}
\ee
and
\bea
\hbar \frac{\partial \phi_p}{\partial t}= - \frac{\delta {\cal H}}{\delta n_p} =-\frac{\partial H}{\partial n_p}=-\mu_p-\frac{\partial {H}^{\rm kin}}{\partial n_p},
\label{josephson_p}
\eea
where
\be
\mu_n=\frac{\partial E(n_n, n_p)}{\partial n_n}\,\,\,{\rm and}\,\,\, \mu_p=\frac{\partial E(n_n, n_p)}{\partial n_p}
\ee
are the neutron and proton chemical potentials when the phases of the condensates do not vary in space.\footnote{From the discussion after Eq.~(\ref{Schroedinger}) it follows that the derivative $\partial E/\partial n_n$ must be regarded as the limit for small integer $\nu$ of \mbox{$[{\mathcal E}(N_n, N_p)-{\mathcal E}(N_n-2\nu, N_p)]/2\nu V$}, where $V$ is the volume of the system.  Similar results apply for the proton chemical potential.  Consequently,  odd--even effects due to pair breaking do not enter in the derivatives.} {Since we consider matter that is electrically neutral, the quantity $\mu_p$ is the energy to add an electron and a proton but, for notational simplicity, we do not indicate this explicitly.}

Quite generally, the equations of continuity for neutrons and for protons have the form
\be
\frac{\partial n_n}{\partial t}+\bm{\nabla}\cdot {\bf j}_n=0.
\label{cont_n}
\ee
and
\be
\frac{\partial n_p}{\partial t}+\bm{\nabla}\cdot {\bf j}_p=0.
\label{cont_p}
\ee

The neutron density and the phase of the neutron condensate  are conjugate variables, and these results also follow from the Hamilton equation, $\partial n_n/\partial t=  \delta {\cal H}/\delta \phi_n$, with the expression for the current given in Eq. (\ref{current_n}).\footnote{Strictly speaking, the conjugate variables are $\hbar \phi_\alpha$ and $n_\alpha$ but we shall generally work in units in which $\hbar$ is equal to unity.}  Similar results hold for the protons.
In the Hamiltonian formalism, the ``coordinates'' and ``momenta'' are to be regarded as independent variables.  Consequently, the derivatives on the right hand sides of  Eqs.\ (\ref{josephson_n}) and (\ref{josephson_p}) are to be evaluated at fixed ${\bf p}_n$ and ${\bf p}_p$.

The basic thermodynamic identity at zero temperature may thus be written as
\be
dE^{\rm tot}= \mu_n^{\rm tot}d n_n+\mu_p^{\rm tot}d n_p + {\bf j}_n \cdot d {\bf p}_n+ {\bf j}_p \cdot d {\bf p}_p,
\ee
where the energy density $E^{\rm tot}$ and the chemical potentials  $\mu_\alpha^{\rm tot}$ all include  kinetic contributions.

The velocities of the components are defined by
 \be
 {\bf v}_\alpha=\frac{{\bf j}_\alpha}{n_\alpha},
 \label{v_alpha}
 \ee
 and therefore it follows from Eqs.~(\ref{current_n}) and (\ref{current_p}) that
 \be
 {\bf v}_\alpha=\frac1{n_\alpha m}\sum_{\beta}n_{\alpha\beta}{\bf p}_\beta.
 \ee

 For a Galilean-invariant system, there are relationships between the $n_{\alpha\beta}$ {\citep{Mendell1991,BorumandJoyntKluzniak1996}}.  Under a transformation to a frame moving with respect to the original frame by a velocity $-\bf v$, the phases $\phi_\alpha$ are increased by an amount $m{\bf v}\cdot{\bf r}/\hbar$.  Consequently, the current density of neutrons is increased by an amount $(n_{nn}+n_{np}) {\bf v}$.  However, from Galilean invariance, the change in the neutron current density is $n_n{\bf v}$ and therefore
 \be
 n_{nn}+n_{np}=n_n.
 \label{Gal_n}
 \ee
 Similarly, by considering the proton current density one finds
 \be
 n_{pp}+n_{np}=n_p.
 \label{Gal_p}
 \ee
 Therefore Eq.\ (\ref{H^kin2}) may be written as
 \be
 {H}^{\rm kin}=\frac{n_n}{2m}p_n^2+\frac{n_p}{2m} p_p^2 -\frac{n_{np}}{2m}({\bf p}_n-{\bf p}_p)^2.
\label{H^kin3}
\ee

 \section{Euler equations}
 The generalizations of Euler's equation for a single component fluid to the two-fluid case are obtained by taking the gradient of Eqs.~(\ref{josephson_n}) and (\ref{josephson_p}) and have the form
\be
\frac{\partial {\bf p}_n}{\partial t}=-\bm{\nabla}\left(\mu_n+\frac{p_n^2}{2m} -\frac1{2m}\frac{\partial n_{np}}{\partial n_n}({\bf p}_n-{\bf p}_p)^2\right)
\label{euler_n}
\ee
and
 \be
\frac{\partial {\bf p}_p}{\partial t}= - \bm{\nabla}\left(\mu_p+\frac{p_p^2}{2m} -\frac1{2m}\frac{\partial n_{np}}{\partial n_p}({\bf p}_n-{\bf p}_p)^2\right),
\label{euler_p}
\ee
 since ${\bf p}_\alpha =\bm{\nabla}\phi_\alpha$.
We may write the terms nonlinear in the ${\bf p}_\alpha$ in Eqs.~ (\ref{euler_n}) and (\ref{euler_p}) by using the vector identity $\bm{\nabla} ({\bf a}^2/2)= {\bf a}\bm\cdot\bm{\nabla}~ {\bf a}+{\bf a}\bm{\times}\bm{\nabla} \bm{\times} {\bf  a}$.  Since in this article we shall consider only situations in which there are no singularities in the flow, we may put $\bm{\bm{\nabla}} \bm{\times} {\bf p}_\alpha=0$ everywhere, and therefore
 \begin{align}
\frac{\partial {\bf p}_n}{\partial t}&+\frac1m {\bf p}_n\cdot \bm{\nabla}   {\bf p}_n     -   \frac1m\frac{\partial n_{np}}{\partial n_n}({\bf p}_n-{\bf p}_p)\bm{\cdot}\bm{\nabla}({\bf p}_n-{\bf p}_p) \nonumber \\ &=-\bm{\nabla}\mu_n +\frac1{2m}\left(\bm{\nabla} \frac{\partial n_{np}}{\partial n_n}\right)({\bf p}_n-{\bf p}_p)^2
\label{euler_n2}
\end{align}
and

 \begin{align}
\frac{\partial {\bf p}_p}{\partial t}&+\frac1m {\bf p}_p\cdot \bm{\nabla}   {\bf p}_p  -   \frac1m\frac{\partial n_{np}}{\partial n_p}({\bf p}_n-{\bf p}_p)\bm{\cdot}\bm{\nabla}({\bf p}_n-{\bf p}_p) \nonumber \\ &=-\bm{\nabla}\mu_p +\frac1{2m}\left(\bm{\nabla} \frac{\partial n_{np}}{\partial n_p}\right)({\bf p}_n-{\bf p}_p)^2  .
\label{euler_p2}
\end{align}
An interesting point is that the additional contribution to the nonlinear terms in the generalization of Euler's equation is proportional to derivatives of $n_{np}$, as  feature already present in the work of \citet[Eqs.~14, 15, 29 and 30]{Mendell1991}.

Equations (\ref{euler_n2}) and (\ref{euler_p2}) may be expressed in terms of the velocities of the components, but the resulting equations are lengthy because of the numerous places where density derivatives of $n_{np}$ appear. {In the Appendix we show that the Euler-like equations in some earlier work do not agree with Eqs.~(\ref{euler_n}) and (\ref{euler_p}).}

\section{Collective modes}
\subsection{Linear modes}

We first consider the frequencies of modes corresponding to small deviations from the situation in which both superfluids are at rest (${\bf p}_n={\bf p}_p=0$).
For a disturbance $\sim {\rm e}^{{\rm i}(\bm{k\cdot r}-\omega t)}$,  the  perturbations  of $\bm{p}_\alpha$ must be in the direction of $\bm k$ and Eqs.~(\ref{cont_n}), (\ref{cont_p}), (\ref{euler_n2}) and (\ref{euler_p2}) when linearized may be written in the matrix form
%\begin{widetext}
\begin{equation}
\label{matrix_formula_linear}
\left(
                     \begin{array}{cccc}
                       -mv              &  0 & n_{nn}       & n_{np} \\
                              0  & -mv     & n_{pn}       & n_{pp} \\
                        {E_{nn}} & E_{np}&   -v  &0  \\
                       E_{np}&E_{pp} & 0 & -v \\
                     \end{array}
                   \right)
                   \begingroup
\renewcommand*{\arraystretch}{1}
                   \left(\begin{array}{c}
  \delta n_n \\
  \delta n_p \\
  \delta p_n \\
  \delta p_p
\end{array}\right)
\endgroup
=0,
\end{equation}
where $v=\omega/k$ is the phase velocity of the wave.
%\end{widetext}
The mode frequencies are determined from the zeros of the determinant of the matrix, i.e.,
\be
v^4- c_s^2v^2+\frac{1}{m^2}\det[E_{\alpha\beta}]\det[n_{\alpha\beta}]=0,
\ee
or

\begin{equation}
\label{omega_coup}
v_\pm^2=\frac{c_s^2}{2} \pm\sqrt{\left(\frac{c_s^2}{2}\right)^2-\frac{\det[E_{\alpha\beta}]\det[n_{\alpha\beta}]}{m^2}},
\end{equation}
where
\begin{equation}\label{c_s}
c_s^2=\frac{E_{pp}n_{pp}+2E_{np}n_{np}+E_{nn}n_{nn}}{m}.
\end{equation}
Equation (\ref{omega_coup}) is a generalization of the result of \citet{BedaqueReddy2014} to allow for entrainment.
In the absence of coupling between neutrons and the charged particles ($E_{np}=0$, $n_{np}=0$, $n_{nn}=n_n$, and $n_{pp}=n_p$), the mode velocities are given by
\be
v_n^0=\left(\frac{n_n E_{nn}}{m}\right)^{1/2}
\label{vn0}
\ee
for the neutrons and
\be
v_p^0=\left(\frac{n_p E_{pp}}{m}\right)^{1/2}
\label{vp0}
\ee
for the charged particles.

One sees from Eq.~(\ref{omega_coup}) that mode frequencies become imaginary if $\det[E_{\alpha\beta}]$ or $\det[n_{\alpha\beta}]$ become negative.  The first condition corresponds to an instability to formation of a density wave with proton and neutron densities in phase for $E_{np}<0$ and out of phase for $E_{np}>0$.  Generalizations of this result to finite wavelengths have previously been employed to obtain estimates of the density at which the transition between uniform matter at higher densities and inhomogeneous matter in the crust occurs  \citep{BBP1971, Hebeler2013}.   The condition $\det[n_{\alpha\beta}]<0$ signals an instability to counterflow of the two components but,  as we shall see in Section \ref{Sec:Entrainment}, this is not expected to occur in the outer core of a neutron star. \\

\subsection{Two-stream instability}
 We now consider small perturbations about a state in which the densities are uniform and the gradients of the phases are also uniform with values ${\bf p}^0_n$ and ${\bf p}^0_p$.  On linearizing Eqs. (\ref{cont_n}), (\ref{cont_p}), (\ref{euler_n3}) and (\ref{euler_p3}), one finds
  \bea
 \frac{\partial n_n}{\partial t}+\frac{n_{nn}}{m}\bm{\nabla}\bm{\cdot} \delta {\bf p}_n+\frac{n_{np}}{m} \bm{\nabla}\bm{\cdot}\delta {\bf p}_p \nonumber \\+\frac{{\bf p}^0_n}{m} \bm{\cdot\nabla} n_{nn}+\frac{{\bf p}^0_p}{m} \bm{\cdot\nabla}n_{np}=0,
 \label{cont_n2}
 \eea
 \vspace{-1em}
  \bea
 \frac{\partial n_p}{\partial t}+\frac{n_{pp}}{m}\bm{\nabla}\bm{\cdot} \delta {\bf p}_p+\frac{n_{np}}{m} \bm{\nabla}\bm{\cdot}\delta {\bf p}_n \nonumber \\+\frac{{\bf p}^0_p}{m} \bm{\cdot\nabla} n_{pp}+\frac{{\bf p}^0_n}{m} \bm{\cdot\nabla}n_{np}=0,
  \label{cont_p2}
 \eea
  \vspace{-1em}
 \bea
 \frac{\partial {\delta}{\bf p}_n}{\partial t}+\frac{{\bf p}^0_n}{m}  \bm{\cdot\nabla} \delta {\bf p}_n  -   \frac1m\frac{\partial n_{np}}{\partial n_n}({\bf p}^0_n-{\bf p}^0_p)\bm{\cdot\nabla}(\delta {\bf p}_n-&\delta{\bf p}_p) \nonumber\\      +\bm{\nabla}\left(\mu_n-\frac1{2m} \frac{\partial n_{np}}{\partial n_n}({\bf p}^0_n-{\bf p}^0_p)^2\right)=0,
 \label{euler_n3}
 \eea
  \vspace{-1em}
 \bea
 \frac{\partial {\delta}{\bf p}_p}{\partial t}+\frac{{\bf p}^0_p}{m}  \bm{\cdot\nabla} \delta {\bf p}_p  -   \frac1m\frac{\partial n_{np}}{\partial n_p}({\bf p}^0_n-{\bf p}^0_p)\bm{\cdot\nabla}(\delta {\bf p}_n-&\delta{\bf p}_p) \nonumber\\      +\bm{\nabla}\left(\mu_p-\frac1{2m} \frac{\partial n_{np}}{\partial n_p}({\bf p}^0_n-{\bf p}^0_p)^2\right)=0,
 \label{euler_p3}
\eea
where $\delta {\bf p}_n={\bf p}_n-{\bf p}^0_n$ and $\delta {\bf p}_p={\bf p}_p -{\bf p}^0_p$. On physical grounds one expects the most unstable mode to be one in which the wave number, and therefore also the velocity perturbations,  are parallel to $\Delta {\bf p}={\bf p}^0_n-{\bf p}^0_p$.  For that case, Eqs.~(\ref{cont_n2}), (\ref{cont_p2}), (\ref{euler_n3}) and (\ref{euler_p3}) may be written in the matrix form
  \begin{widetext}
\begin{equation}
\label{matrix_formula2}
\left(
                     \begin{array}{cccc}
                       -\frac{m\omega }{k}+p_n^0- \Delta p\frac{\partial n_{np}}{\partial n_n}               & -\Delta p\frac{\partial n_{np}}{\partial n_p}     & n_{pp}       & n_{np} \\
                                -\Delta p\frac{\partial n_{np}}{\partial n_n}& -\frac{m\omega }{k}+p_p^0- \Delta p\frac{\partial n_{np}}{\partial n_p}            & n_{pn}       & n_{nn} \\
                        E_{nn}-\frac1{2m}\frac{\partial^2 n_{np}}{\partial n_n^2}(\Delta p)^{2} & E_{np}-\frac1{2m}\frac{\partial^2 n_{np}}{\partial n_n \partial n_p}(\Delta p)^2&-\frac{\omega}{k} +\frac{p_n^0}{m}-\frac{\partial n_{np}}{\partial n_n} \frac{\Delta p}{m} & \frac{\partial n_{np}}{\partial n_n}\frac{\Delta p}{m}\\
                       E_{np}-\frac1{2m}\frac{\partial^2 n_{np}}{\partial n_n \partial n_p}(\Delta p)^2&E_{pp}-\frac1{2m}\frac{\partial^2 n_{np}}{\partial n_p^2}(\Delta p)^2 &
                        \frac{\partial n_{np}}{\partial n_p}\frac{\Delta p}{m}& -\frac{\omega}{k} +\frac{p_p^0}{m}-\frac{\partial n_{np}}{\partial n_p} \frac{\Delta p}{m}\\
                   \end{array}
                   \right)
                   \begingroup
\renewcommand*{\arraystretch}{1.7}
                   \left(\begin{array}{c}
  \delta n_n \\
  \delta n_p \\
  \delta p_n \\
  \delta p_p
\end{array}\right)
\endgroup
=0.
\end{equation}
\end{widetext}
The mode frequencies are given by the condition that the determinant of the matrix in Eq.\ (\ref{matrix_formula2}) vanish.  This result is a generalization to allow for entrainment of the results of \citet{AnderssonComerPrix2004}.  Equation (\ref{matrix_formula2}) illustrates that fact that, in nonlinear problems,  density derivatives of $n_{np}$ occur, as well as the quantity itself.

\section{Applications to the outer core}
\label{Unif}
\subsection{Equation of state}
The equation of state that we use is based on chiral effective field theory (EFT), in which the symmetries associated with QCD are built into an effective Hamiltonian for nucleons   \citep{EpelbaumHammer2009}.The parameters of the theory are determined from nucleon--nucleon scattering and other low-energy nuclear data.   The particular version of the theory that we shall use is that of \citet{Hebeler2013}, in which an analytic fit is made to calculations for pure neutron matter and symmetric nuclear matter and an interpolation is made for proton fractions $x=n_p/n$ intermediate between the two proton fractions $x=0$ and $x=1/2$ for which microscopic calculations have been made. Here
\begin{equation}\label{n_def}
n=n_p+n_n.
\end{equation}
is the total density of nucleons.  The nuclear part of the energy per particle (without electrons) is given by \citet{Hebeler2013}
\begin{eqnarray}
&&\frac{\epsilon}{\epsilon_0} = \frac{3}{5}\left[ {{x}^{5/3}}+{{\left( 1-x \right)}^{5/3}} \right]{{\left(
\frac{2}{{{n}_{0}}} \right)}^{2/3~}}{{n}^{2/3}}\\
\label{EOS}
&&-\left[ {{\alpha }_{1}}x\left( 1-x \right)+{{\alpha }_{2}} \right]\frac{n}{{{n}_{0}}}+\gamma\left[ {{\eta }_{1}}x\left(
1-x \right)+{{\eta }_{2}} \right]{{\left( \frac{n}{{{n}_{0}}} \right)}^{\gamma }},\nonumber
\end{eqnarray}
 and the values of the parameters
are ${\epsilon }_{0}=36.84\:\mathrm{MeV}$, and $\alpha_1=6.14$, $\eta_1=4.02$,
$\alpha_2=1.4$, and $\eta_2=0.9$.

This form is expected to be a reasonable approximation for baryon densities  in the range $\sim 0.08-0.2\,\mathrm{fm}^{-3}$.
The energy density is the sum of the nucleon energy density and the electron contribution
\begin{equation}\label{Eeps}
E=n\varepsilon+E_e.
\end{equation}
In the formalism described above, it is assumed that the number of neutrons and the number of protons are conserved.  This is a good approximation when the time scales of interest in the motions are short compared with the time scale for weak interactions.  We have made no assumption about the ratio of neutrons to protons, but in the numerical calculations we shall concentrate on the case of matter in beta equilibrium, which should be a good first approximation for most of the life of a neutron star.   The  condition for beta equilibrium is that $m_pc^2+\partial E/\partial n_p+\mu_e=m_nc^2+\partial E/\partial n_n$ \citep{BBP1971} which, with the neglect of the difference between the neutron and proton masses, gives
\begin{equation}\label{eqX}
 \partial\epsilon/\partial x+\mu_e\approx0,
\end{equation}
where $\mu_e=\partial E_e/\partial{n_e}$ is the electron chemical potential, which for ultrarelativistic degenerate electrons is
\begin{equation}\label{mue}
  \mu_e=\hbar c(3\pi^2n_e)^{1/3}.
\end{equation}
Bulk matter is electrically neutral and therefore
\begin{equation}\label{electr_neutr}
n_e=n_p.
\end{equation}
The equilibrium value of the proton fraction calculated from Equations (\ref{EOS}) and (\ref{eqX}) is shown in Figure
1(a).
\begin{figure}
\centering
\includegraphics[width=3.5in]{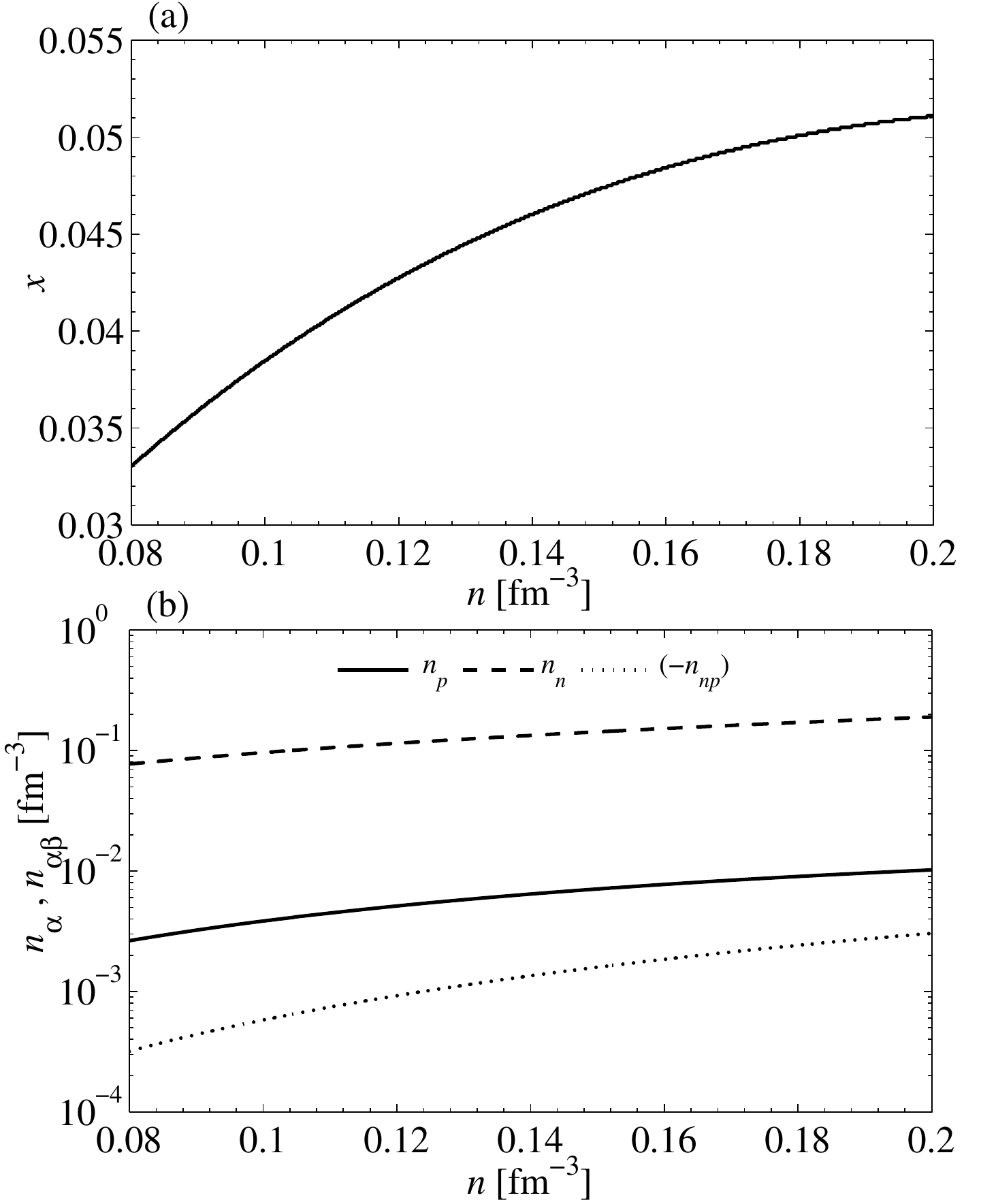}
\caption{Equilibrium proton fraction and nucleon number densities calculated from the equation of state of \citet{Hebeler2013}.  (a) Proton fraction in beta equilibrium as a function of nucleon number density, calculated from
Equations (\ref{EOS}) and (\ref{eqX}). (b) Equilibrium values of the nucleon densities $n_n$ and $n_p$. Also shown are results for $n_{np}$  calculated from Landau Fermi-liquid theory with the Skyrme interaction SLy4 (see Eq.~(\ref{n_npSkyrme}).}
\end{figure}
For convenience, nucleon densities for matter in beta equilibrium are
plotted as functions of baryon density in Figure~1(b).

\newpage

\subsection{Thermodynamic derivatives}
\label{Thermo}
The second derivatives of the total energy density $E$,
\begin{equation}\label{Eab}
E_{\alpha\beta}=\frac{\partial^2E}{\partial n_\alpha\partial n_\beta},
\end{equation}
determine observable properties such as sound speeds.
From Eq.\ (\ref{Eeps}) it follows that
\begin{eqnarray}
\label{Epp}
  &&E_{pp}=\frac{\partial^2\left(n\epsilon\right)}{\partial n_p^2} + \frac{\partial\mu_e}{\partial
  n_e},\\
\label{Enn}
  &&E_{nn}=\frac{\partial^2\left(n\epsilon\right)}{\partial n_n^2}, \\
\label{Enp}
  &&E_{np}=\frac{\partial^2\left(n\epsilon\right)}{\partial n_p\partial n_n}.
\end{eqnarray}
We express derivatives with respect to particle density in terms of the variables $n$ and $x$ by using the relationships
\begin{eqnarray}
\label{d_dnn}
\left.\frac{\partial }{\partial n_n}\right|_{n_p}=-\frac{x}{n}\frac{\partial}{\partial x}+\frac{\partial}{\partial n},\,\,\,{\rm and}\,\,\,\\
\label{d_dnp}
\left.\frac{\partial }{\partial n_p}\right|_{n_n}=\frac{1-x}{n}\frac{\partial}{\partial x}+\frac{\partial}{\partial n}.
\end{eqnarray}
 The results for the derivatives $E_{\alpha\beta}$ obtained from Eqs.~(\ref{Epp})--(\ref{Enp})  and (\ref{EOS}) are plotted in Fig.~2.  The quantity $E_{pp}$ has contributions from both protons and electrons, and we show the difference between $E_{pp}$ and the contribution from electrons, which in the absence of screening is $\partial \mu_e/\partial n_e$. One sees that the electronic contribution to $E_{pp}$ is dominant.

\begin{figure}
\includegraphics[width=3in]{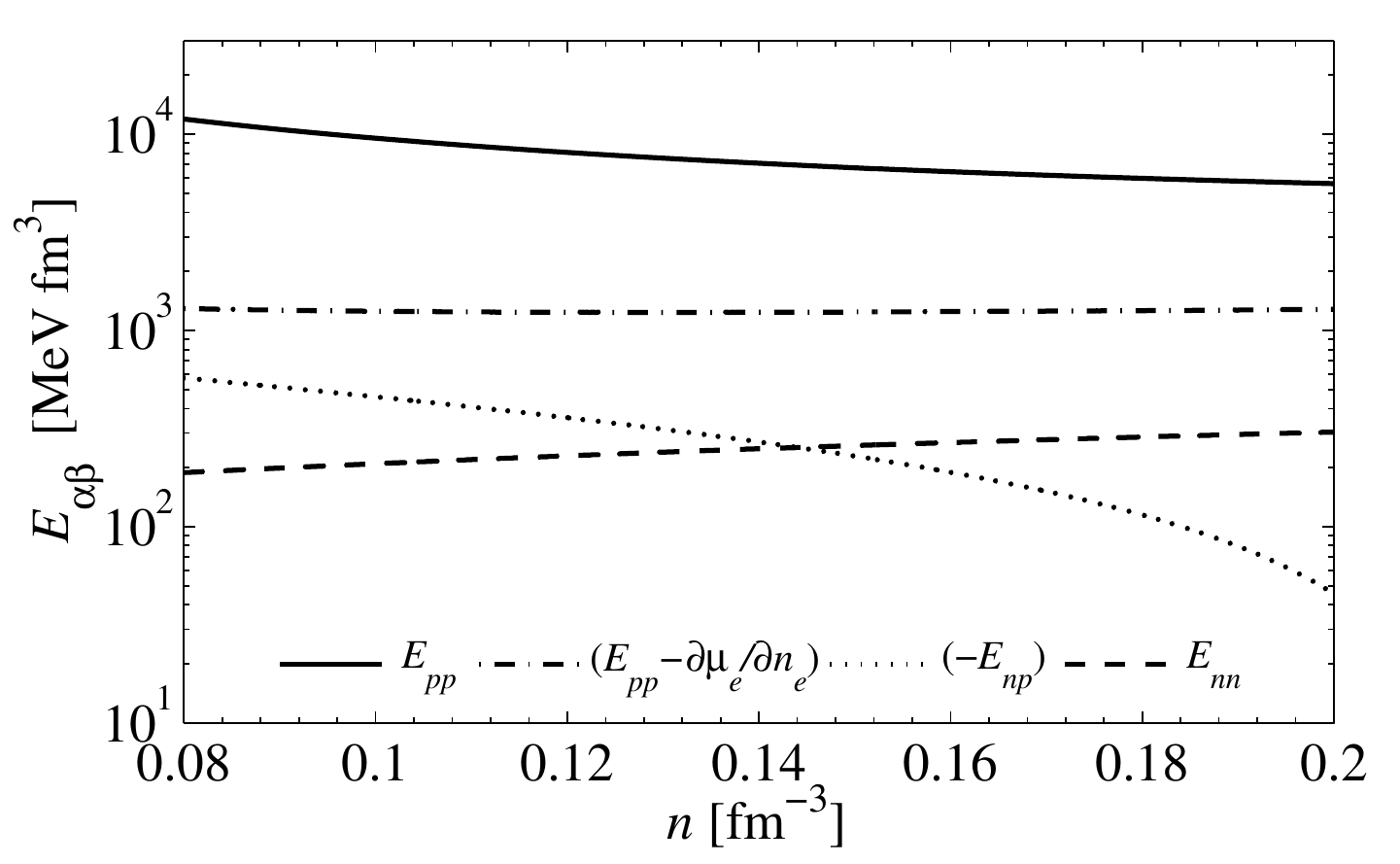}
\caption{Thermodynamic derivatives $E_{\alpha\beta}$ and $\partial \mu_e/\partial n_e$, the electronic contribution to $E_{pp}$, for baryon densities  in the  outer core. The
equation of state is taken from \citet{Hebeler2013}.}
\end{figure}

\subsection{Entrainment}
\label{Sec:Entrainment}

In addition to interactions between the densities of the various components, there are also interactions between the flows of the two components, which are reflected in non-zero values of $n_{np}$, an effect often referred to as entrainment.   In the outer core of neutron stars, pairing gaps are expected to be of order 1 MeV or less, while nucleon Fermi energies are one or two orders of magnitude larger.  Thus pairing contributes little to the total energy, and one may use Landau's theory of normal Fermi liquids to calculate $n_{np}$ and  \citet{BorumandJoyntKluzniak1996} find
\begin{equation}
\frac{n_{np}}m=\frac{ k_{Fp}^{2}k_{Fn}^{2}}{9\pi^4}f_1^{np},
\label{rho_np_FL}
\end{equation}
where $k_{F\alpha}$ are the Fermi wave numbers of neutrons and protons, and $f_1^{np}$ is the $l=1$ component of the
Landau parameter for the interaction between neutrons and protons.  A general treatment of entrainment at nonzero temperature has been given by \citet{GusakovHaensel}.

Most microscopic calculations of Landau parameters for nuclear matter have been performed for either symmetric nuclear matter or for pure neutron matter (For recent examples see, e.g., \citet{LombardoEtal2011,Holt2013}),   and there is a need for further study of matter with proton fractions $\sim 5\%$ of interest for neutron star cores.  An exception is the work of \citet{ChamelHaensel2006}, who gave a general treatment of entrainment and made specific calculations for effective interactions of the Skyrme type.
For the standard form of the Skyrme interaction \cite[Eq. (23)]{ChamelHaensel2006}, the entrainment comes solely from the terms involving gradients of the wave function and by direct calculation one finds
\be
f_1^{np}=-\frac{k_{Fn}k_{Fp}}2 \left[t_1\left(1+\frac12x_1\right) +t_2\left(1+\frac12 x_2\right)\right]
\label{f1Skyrme}
\ee
and therefore, from Eqs.~(\ref{rho_np_FL}) and (\ref{f1Skyrme}),
\begin{equation}\label{rho12}
  n_{np}=\tilde{\alpha}_{np}n_n n_p
\end{equation}
in the notation of \citet{ChamelHaensel2006}, with
\be
\tilde{\alpha}_{np}= -\frac{m}2{}    \left[t_1\left(1+\frac12x_1\right) +t_2\left(1+\frac12 x_2\right)\right].
\ee
For the Skyrme interaction SLy4 developed especially for astrophysical applications, $t_1=486.82$ MeV fm$^5$,  $x_1=-0.344$, $t_2=-546.39$ MeV fm$^5$, and $x_2=-1.000$ \citep{Chabanat_et_al} and therefore
\be
n_{np}\approx -1.567 \,{\rm fm}^3 n_n n_p.
\label{n_npSkyrme}
\ee
As \citet{ChamelHaensel2006} showed, the 27 Skyrme interactions recommended for astrophysical applications by \cite{Stone_et_al} give values for $\tilde{\alpha}_{np}$ between 0 and \mbox{$-10.4$ fm$^3$}, while the Skyrme interactions developed by the Lyon group lead to values of around $-1.5$ fm$^3$, with the exception of SLy230a, for which it is essentially zero.   The wide range of values of $n_{np}$ that Skyrme interactions predict underscores the need to pin down its value better from more fundamental considerations.

The conditions for stability to counterflow of the two components are that $n_{nn}$, $n_{pp}$, and $\det[n_{\alpha\beta}]$ are positive.  If the third condition and one of the first two are satisfied, the remaining condition holds automatically.   For the Skyrme interactions that have been considered above, $n_{np}$ is negative and therefore from Eqs.~(\ref{Gal_n}) and (\ref{Gal_p}) it follows that the first two conditions hold.  Since
\be
\det[n_{\alpha\beta}]=(n_n  -n_{np})(n_p-n_{np})-n_{np}^2=n_n n_p-n_{np} n>0,
\ee
the third condition also holds and matter is stable to counterflow.
\begin{figure}
\includegraphics[width=3in]{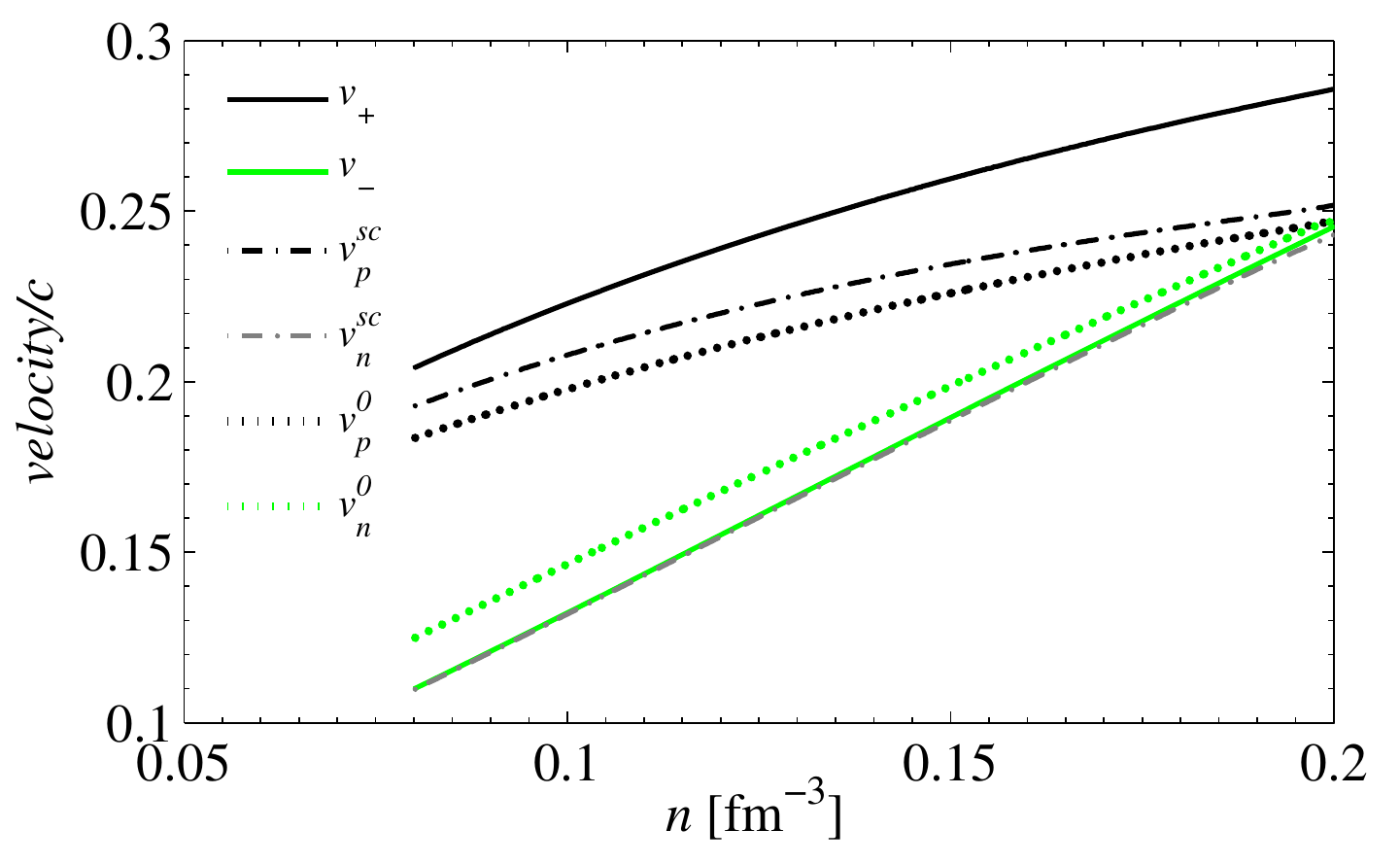}
\caption{Sound speeds $v$ in the absence of counterflow in units of $c$ obtained from Equation~(\ref{omega_coup}),
as functions of the baryon number density  (solid lines).
The dotted lines correspond lines correspond to the velocities in the absence of coupling between neutrons and protons, Equation (\ref{vn0}) (lower curve) and Equation (\ref{vp0}) (upper curve).  The dot--dashed lines ($v^{sc}_\pm$) show the results in the absence of entrainment ($ \det n_{\alpha\beta}=n_nn_p$).  The modes corresponding to the three uppermost curves are dominated by motion of charged particles, while motion of neutrons is predominant in the modes corresponding to the three lowermost curves.}
\end{figure}

\subsection{Collective mode frequencies}
\label{Modes}
In Figure 3 we show results for the velocities of longitudinal collective modes.  The velocities of modes in the absence of coupling between neutrons and protons are given by Eqs.\ (\ref{vn0}) and (\ref{vp0}) and the thermodynamic derivatives are taken from Sec.~\ref{Thermo}.  The dashed lines $v^{sc}_\pm$ include the effects of $E_{np}$ but the effects of entrainment are neglected ($n_{nn}=n_n$, $n_{pp}=n_p$, and $n_{np}=0$).  Finally,  the full lines include both the effects of nonzero $E_{np}$ and entrainment, Eq.~(\ref{omega_coup}).  Entrainment affects the charged particle mode more than the neutron one since $n_{np}/n_p$ is more than an order of magnitude larger than $n_{np}/n_n$.  The hybridization of the charged particle and neutron modes is relatively weak.  When $E_{np}$ is nonzero but the effects of entrainment are neglected,
the velocity of the charged particle mode is raised, while that of the neutrons is lowered.   Entrainment has little effect on the velocity of the neutron mode but further raises that of the charged-particle mode.

It is instructive to compare properties of the outer core with those of the inner crust, where the protons reside in nuclei.  For the inner crust, values are taken from \citet{KobyakovPethick2016}, which corrected a coding error in the paper of \citet{KobyakovPethick2013}.  These were based on the equation of state of \citet{LattimerSwesty}. In Figure 4 we plot  values
of the thermodynamic derivatives $E_{\alpha\beta}$ for the inner crust and the outer core.  Despite the different equations of state in the crust and the core regions, the values of $E_{\alpha\beta}$ are rather similar at the crust--core boundary.

\begin{figure}
\includegraphics[width=3in]{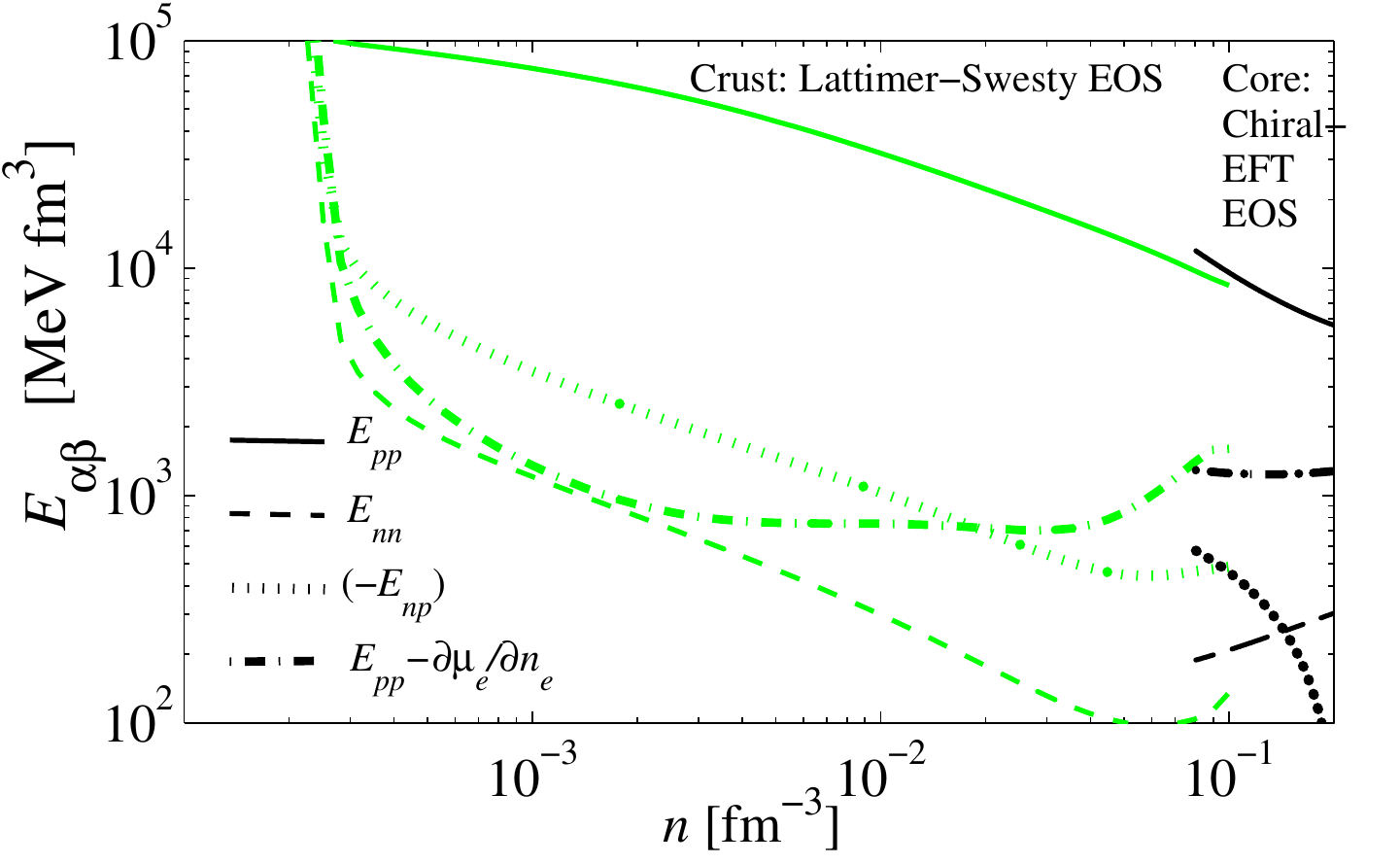}
\caption{Thermodynamic derivatives $E_{\alpha\beta}$ in the inner crust and outer core. The equations of state (EOS) used are indicated in the figure. }
\end{figure}

Sound speeds across the crust--core transition region in the star are shown in Figure 5.
It is interesting to note that, while the
$E_{\alpha\beta}$ are almost continuous between the crust and the core,
the sound speeds exhibit significant jumps.
\begin{figure}
\includegraphics[width=3in]{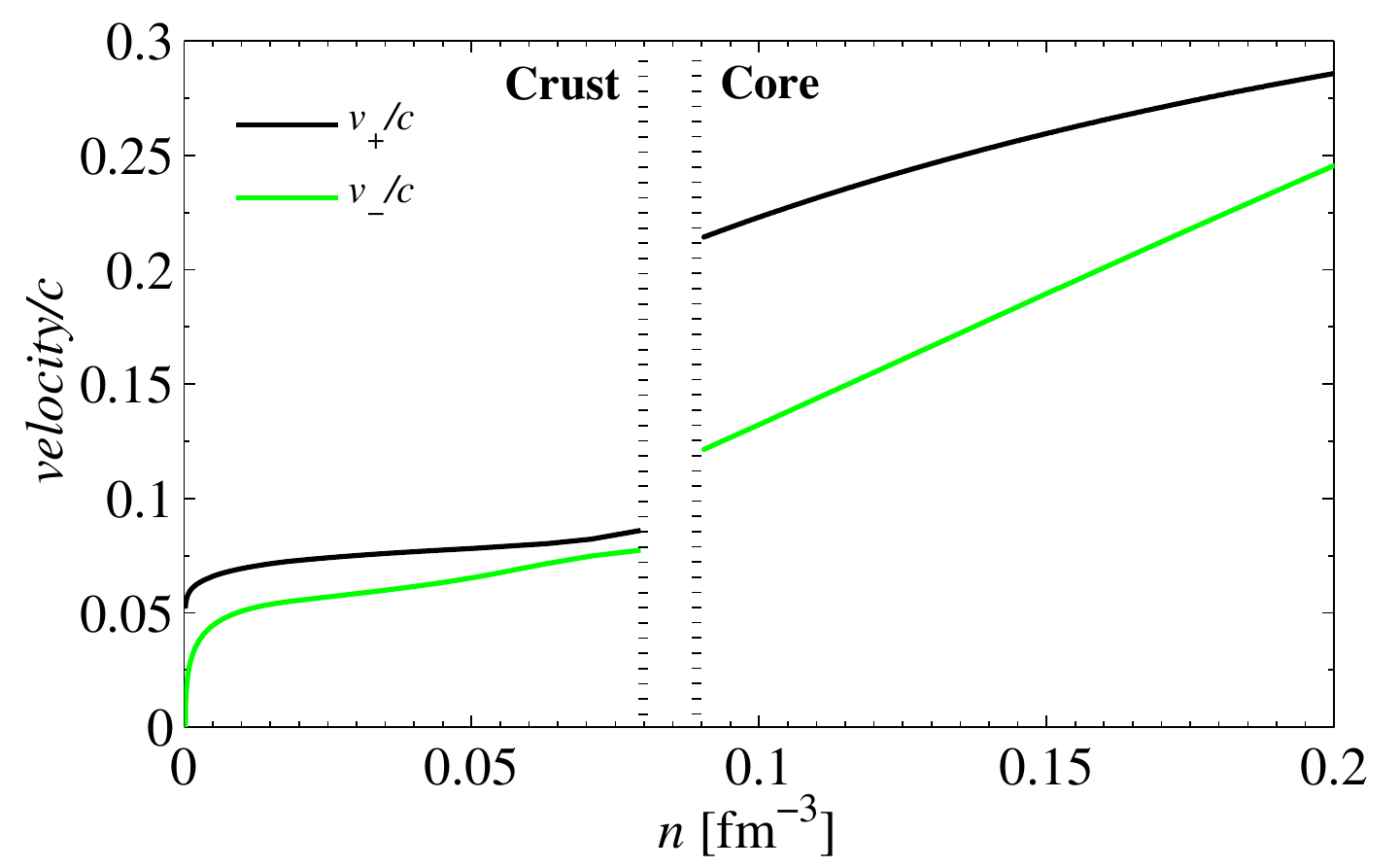}
\caption{Speed of longitudinal sound-like modes in the inner crust and outer core of a neutron star as a function of baryon density. For the outer core, the results correspond to those in  Figure 4. In the calculations for the inner crust, the neutron superfluid density was taken to be the density of neutrons outside nuclei.
For the crust, the results are those of \citet{KobyakovPethick2016}.
The neutron mode (green line) tends to zero at the neutron drip density $n_{ND}\approx2.2\times10^{-4}\,\mathrm{fm}^{-3}$. The left vertical dotted line corresponds to the maximum density for which the assumption of spherical nuclei in the Lattimer--Swesty model is still valid \citep{KobyakovPethick2013}. The right vertical line corresponds to the lower bound of density of uniform nuclear matter, below which it is unstable to formation of a density wave \citep{Hebeler2013}.}
\end{figure}
At the boundary, the charged-particle mode is about three times slower in the crust than in the core; this is due to the fact that entrainment in the crust is much greater than in the core by about one order of magnitude because in nuclei the number of neutrons entrained by a single proton is of order the neutron to proton ratio in nuclei, $\sim 10$, at the inner boundary of the crust.

\section{Discussion}
\label{Conclusion}

In this paper we have generalized to a two-component fluid Euler's equation for a  single component.  The approach we have adopted is based on the Josephson equation for the phases of the condensate wave functions of the nucleons and the continuity equations.  This makes possible a direct derivation of the basic results.
The nonlinear terms in the Euler-like equations have contributions proportional to density derivatives of the strength of the entrainment.   These contributions do not affect small oscillations about a state in which the two fluids are at rest, but they do enter in, e.g., the condition for the two-stream instability.   These terms are implicit in the work of \citet{Mendell1991}  and they arise from the effects of entrainment on the nucleon chemical potentials.\footnote{The original work of \citet{AndreevBashkin1975} on entrainment in the helium liquids did not mention explicitly the entrainment contributions to the chemical potentials.  However, this had no influence on the applications described in that paper, which were to linear modes.}

In some earlier treatments,  the energy due to entrainment was regarded as part of an ``internal energy'' defined as the difference between the total energy and the kinetic energy in the absence of entrainment  \citep{Prix2004, AnderssonComerPrix2004}, but  the approach presented here shows that it is natural to treat the energy due to entrainment as part of the ``kinetic energy''.  In this way it is made clear that the nucleon chemical potentials contain contributions proportional to derivatives of the entrainment energy density with respect to the neutron and proton densities.  The thermodynamic potential appropriate when the system is specified by the number densities of neutrons and protons and the phases of the condensates is the Hamiltonian, that is the total energy of the system, while its Legendre transform,
\bea
\Xi&=& {\bf p}_n \cdot {\bf j}_n +{\bf p}_p \cdot {\bf j}_p   -E^{\rm tot} \nonumber \\
 &=&\frac12\sum_{\alpha\beta}     m ({\mathsf n}^{-1})_{\alpha\beta} ~ {\bf j}_\alpha \cdot {\bf j}_\beta -E,
 \label{Xi}
\eea
is the potential appropriate when the current densities are regarded as the variables.  Here the matrix ${\mathsf n}^{-1}$ is the inverse of the matrix with elements $n_{\alpha\beta}$. Numerically, the first term on the right side of Eq.\ (\ref{Xi}) is equal to the kinetic energy, Eq.~(\ref{H^kin2}).

Our calculations show that in the generalizations of Euler's equation to two-component superfluid hydrodynamics first and second density derivatives of the entrainment function $n_{np}$ appear.  Nonlinear effects in superfluid hydrodynamics have been investigated in a number of different contexts {\citep{PrixComerAndersson2002,AnderssonComerPrix2004,GusakovAndersson2006,GlampedakisAnderssonSamuelsson2011, Haskell2011, Link2012,PassamontiLander2012}}, and an important task for future work is to
investigate to what extent results are altered by the nonlinear terms derived in the present article. { It is also necessary to reexamine how the terms obtained from a Hamiltonian approach are reflected in the Lagrangian and hybrid approaches used in other work.}

In this article, we have assumed that the flow is irrotational, in the sense that $\bm\nabla \boldsymbol{\times} {\bf p}_n$ and $\bm{\nabla} \bm{\times} {\bf p}_p$ vanish.  We leave for future work the incorporation of electromagnetic fields, vortices, and rotating frames of reference.  An additional direction for  investigation is the effect of nonzero temperature, which results in the appearance of a normal fluid of excitations.

As applications, we have considered oscillations of uniform neutron star matter.  We have generalized the treatment of the two-stream instability given by \citet{AnderssonComerPrix2004}.  To make realistic estimates of the conditions under which the two-stream instability can occur in neutron star cores, it is necessary to take into account damping: in particular, it is important to include pair-breaking processes that will set in at wave numbers of approximately $\Delta/v_F \sim k_F(\Delta/E_F )$, where $\Delta$ is the superfluid gap of a component, $v_F$ its Fermi velocity, $k_F$ its Fermi wave number, and $E_F$ its Fermi energy.  These wave numbers are much less than the respective Fermi wave numbers.

Velocities of sound-like modes in the outer core in the absence of counterflow have been calculated.  In particular, we have generalized the discussion of \citet{BedaqueReddy2014} to allow for entrainment, and we have used recent calculations of the equation of state to evaluate the thermodynamic derivatives.  Extensions of this work to shorter wavelengths and to calculate damping of modes by the electrons will be reported elsewhere \citep{Kobyakov_et_al2016}.

\vspace{2em}

\section*{Acknowledgments}
DK is grateful to Axel Brandenburg, Emil Lundh, Mattias Marklund, Lars Samuelsson and the late Vitaly Bychkov for discussions during the early stages of this work.
We have also enjoyed the hospitality of NORDITA in Stockholm, API in Amsterdam, ISSI in Bern,
ECT* in Trento, and the Niels Bohr Institute in Copenhagen.
This work was supported by the J C Kempe foundation, the Baltic Donation foundation,
by a Nordita Visiting PhD fellowship, by
ERC Grant
307986 Strongint, by the Swedish Research Council (VR) and by the Russian Fund for Basic Research grant 31 16-32-60023/15.

\begin{appendix}
\section{Comparison with earlier work}
The results of the present work agree with the work of \citet{Mendell1991}.
Here we compare these results with those of
other studies that are also based on Mendell's work.
In \citet{AnderssonComer2001} and \citet{AnderssonComerPrix2004} the Euler-like equation for the neutrons has the form
\be
\left(\frac{\partial}{\partial t}+v_{nj}\frac{\partial}{\partial x_j}\right)
[v_{ni}+\epsilon_n(v_{pi}-v_{ni})]
+\frac1{m}\frac{\partial\mu_n^{ACP }}{\partial x_i} +\epsilon_n (v_{pj}-v_{nj})\frac{\partial v_{nj}}{\partial x_i} =0,
\label{A1}
\ee
where the indices $i$ and $j$ refer to Cartesian coordinates, and the index $j=1,2,3$ is to be summed over.
Here
\be
\epsilon_n=\frac{2\alpha}{m n_n}=-\frac{n_p n_{np}}{\det n_{\alpha\beta}}=-\frac{n_{np}}{n_n}\left[ 1-\frac{n_{np}^2}{n_n n_p}       \right]^{-1}
\ee
and we denote by $\mu_n^{ACP}$ the chemical potential used in those papers, which is different from those employed in the present article.
Comparison with the present work is simplified by observing that the combination ${\bf v}_{n}+\epsilon_n({\bf v}_p-{\bf v}_n)$ is what we denote by ${\bf p}_n/m$.  Equation (\ref{A1}) may therefore be written as
\be
\left(\frac{\partial}{\partial t}+\frac{p_{nj}}{m}\frac{\partial}{\partial x_j}\right)
\frac{p_{ni}}{m}
+\frac{\partial}{\partial x_i} \frac{\mu_n^{ACP}}{m}+\epsilon_n (v_{pj}-v_{nj})\frac{\partial}{\partial x_i} v_{nj}    -  \epsilon_n(v_{pj}-v_{nj})\frac{\partial}{\partial x_j} \frac{p_{ni}}{m}=0.
\ee
If the problem is one-dimensional, with all variations in the $x$-direction, one finds
\be
\left(\frac{\partial}{\partial t}+\frac{p_{nx}}{m}\frac{\partial}{\partial x}\right)
\frac{p_{nx}}{m}
+\frac{\partial}{\partial x} \left(\frac{\mu_n^{\rm ACP }}{m}-\frac12\epsilon_n^2 (v_{px}-v_{nx})^2\right)=0.
\label{EulerAC}
\ee
We now contrast this result with the one found from the present work.
Equation (\ref{euler_n}) for the one-dimensional case reads
 \be
\left(\frac{\partial}{\partial t}+\frac{p_{nx}}{m}\frac{\partial}{\partial x}\right)
\frac{p_{nx}}{m}    +
\frac{\partial}{\partial x} \left(\frac{\mu_n}{m}- \frac12
\left[\frac{n_n n_p}{\det n_{\alpha\beta}}\right]^2  \frac{\partial n_{np}}{\partial n_n} (v_{px}-v_{nx})^2\right)    =0,
\label{A6}
\ee
where we have made use of the relation
\be
{\bf v}_p-{\bf v}_n= \frac{ \det n_{\alpha\beta}}{n_n n_p}  \frac{{\bf p}_p-{\bf p}_n}{m},
\label{A5}
\ee
which follows from Eqs. (\ref{current_n}), (\ref{current_p}) and (\ref{v_alpha}).
The  terms containing ${\bf v}_n-{\bf v}_p$  in equations (\ref{EulerAC}) and (\ref{A6}) do not agree.  We have been unable to find in the literature an explicit expression for $\mu_n^{\rm ACP}$.  If it is to be identified with the chemical potential in the absence of flows (what we denote by $\mu_n$), there is a conflict.  There is too if it is identified with $\mu_n-(\partial n_{np}/\partial n_n)({\bf p}_p-{\bf p}_n)^2/2m$, the chemical potential  with the entrainment contribution but not that from the flow of the neutrons.    Similar conclusions apply for the protons.
\end{appendix}
\end{document}